\documentclass{mn2e}
\usepackage{times}
\usepackage{epsf}

\newcommand{\source}{\hbox{3C\,346}}
\newcommand{\asca}{\textit{ASCA}}
\newcommand{\chandra}{\textit{Chandra}}
\newcommand{\rosat}{\textit{ROSAT}}
\newcommand{\einstein}{\textit{Einstein}}

\newcommand{\hst}{\textit{HST}}
\newcommand{\vla}{\textit{VLA}}

\begin{document}

\title[X-ray synchrotron emission from
the jet of \source]{X-ray synchrotron emission from the
oblique shock in the jet of the powerful radio galaxy \source}

\author[D.M. Worrall \& M. Birkinshaw
       ]{D.M. Worrall \& M. Birkinshaw 
\\
Department of Physics, University of Bristol, Tyndall Avenue,
Bristol BS8~1TL}

\maketitle

\label{firstpage}

\begin{abstract}

We report the first detection, with \chandra, of X-ray emission from
the jet of the powerful narrow-line radio galaxy \source.  X-rays are
detected from the bright radio and optical knot at which the jet
apparently bends by about 70 degrees.  The \chandra\ observation also
reveals a bright galaxy-scale atmosphere within the previously-known
cluster, and provides a good X-ray spectrum for the bright core of
\source.  The X-ray emission from the knot is synchrotron radiation,
as seen in lower-power sources.  In common with these
sources, there is evidence of morphological differences between the
radio/optical and X-ray structures, and the spectrum is inconsistent
with a one-component continuous-injection model.  We suggest that the
X-ray-bright knot is associated with a strong oblique shock in a
moderately relativistic, light jet, at $\sim 20$ degrees to the line
of sight, and that this shock is caused by the jet interacting with
the wake in the cluster medium behind \source's companion galaxy.
The general jet curvature can result from pressure gradients in the
cluster atmosphere.
\end{abstract}


\section{Introduction}
\label{sec:intro}

The powerful radio galaxy \source\ is hosted by a 17th magnitude
galaxy at $z = 0.161$ (e.g., Laing, Riley \& Longair 1983).  The
western radio structure shows a terminal hotspot embedded in lobe
emission which displays a Laing-Garrington effect
(Akujor \& Garrington 1995), showing it to lie on the far side
of the nucleus.  By contrast,
the eastern jet (on the observer's side
of the nucleus) is highly distorted (Spencer et al.~1991; Cotton et
al.~1995).  Optical synchrotron emission associated with the eastern
jet was first reported from ground-based observations (Dey \& van
Breugel 1994), and later \hst\ observations show a remarkable
correspondence between optical and radio jet features (de Koff et
al.~1996; de Vries et al.~1997).  In the X-ray, \source\ was first
detected with \einstein\ (Fabbiano et al.~1984), and subsequent
observations with \rosat\ and \asca\ allowed Worrall \& Birkinshaw
(2001) to separate spatially and spectrally the core from X-ray
emission from gas in the potential well of the surrounding cluster,
the existence of which is supported by optical galaxy counts (Zirbel
1997; Harvanek \& Stocke 2002).  Worrall \& Birkinshaw (2001)
predicted that the brightest knot in the radio and optical jet of
\source\ would be detectable with \chandra.  Here we report the
results of an observation made with the principal aim of detecting
this emission and investigating the dynamical state of the radio jet.

In this paper we adopt values for the cosmological parameters
of $H_o =
70$~km s$^{-1}$ Mpc$^{-1}$, $\Omega_{\rm m} = 0.3$, and
$\Omega_\Lambda = 0.7$.  Thus 1~arcsec corresponds to 2.77~kpc at \source. 

\section{Observations}
\label{sec:obs}

\subsection{\chandra}
\label{sec:xrayobs}

We observed \source\ in FAINT data mode with the back-illuminated CCD
chip, S3, of the
Advanced CCD Imaging Spectrometer (ACIS) on board \chandra\ on 2002
August 3.  Details of the instrument and its modes of operation can
be found in the \chandra\ Proposers' Observatory Guide, available from
http://cxc.harvard.edu/proposer.  Results presented here use {\sc ciao
v3.0.2} and the {\sc caldb v2.26} calibration database.  The data have
been re-calibrated and analysed, with random pixelization removed and
afterglow events included,
following the software ``threads'' (http://cxc.harvard.edu/ciao)
from the \chandra\ X-ray Center (CXC).  The time-dependent ACIS gain
correction was applied following the recipe from
http://cxc.harvard.edu/contrib/alexey/tgain/tgain.html.  Only events with
grades 0,2,3,4,6 are used.  

There
were some intervals during the observation when the background rate as
much as doubled, and these periods (about 10\% of the exposure)
were removed, leaving a calibrated
data set with an observation duration of 41.082~ks.

The observation was made with a 256-row subarray, giving a 2 by 8 arcmin
field of view. The subarray was used to reduce the readout time
to 0.84~s and so decrease the incidence of multiple events within
the frame-transfer time. The 0.5--3 keV event rate of roughly 0.06 counts per
frame from the core results in a maximum pile-up over the image
(at the core) of roughly 1 per cent.

\subsection{Radio}
\label{sec:radioobs}

We made a high-resolution (beam-size 0.11 arcsec $\times$ 0.13 arcsec)
image of \source\ from
the \vla-archive A-array U-band (15~GHz) data of observation AO127,
performed in December 1996.  The data were flagged and self-calibrated
using standard procedures within {\sc aips}. Component sizes and
fluxes were measured using the {\sc imfit} task.

\subsection{Optical}
\label{sec:opticalobs}

We have used four imaging data sets from the \hst\ archive to
characterize the morphology of the jet in the optical, and measure the
broad-band optical to ultraviolet spectrum of the X-ray-bright knot.
The data were taken with the STIS camera with (a) the CCD detector and
clear 50CCD filter covering a broad band in the visual with effective
wavelength $\sim 686.6$ nm (data set o61p01010, 828~s exposure, March 2000), (b)
the CCD detector and F28X50LP filter with an effective wavelength
$\sim 764.5$ nm (data set o4al11020, 937~s exposure, October 1997), (c) the NUV-MAMA
detector and F25QTZ filter, with an effective wavelength $\sim 236.4$
nm (data set o5gv06k1q, 1800~s exposure, August 2000), and (d) the FUV-MAMA
detector and F25SRF2 filter with an effective wavelength of $\sim
147.5$ nm (data set o5gv06k6q, 1734~s exposure, August 2000).  The calibrated,
flat-fielded, and distortion-corrected
images were processed with the {\sc iraf} task {\sc
cosmicrays} to remove residual cosmic-ray events.  Net counts in the
X-ray-bright knot were obtained using a box of area $\sim 0.2$ square
arcsec with background measured from a nearby region where the galaxy
surface brightness appeared to be comparable.  The error table
provided with the calibrated data was used to measure the
uncertainties.  The conversion from counts to flux density (in units
of Jy) was obtained using the {\sc synphot} package in {\sc iraf} for
an adopted spectral shape of $S_\nu \propto \nu^{-1.5}$.  To measure
the flux from optically-unresolved knots, such as that
corresponding to the radio knot discussed in section~\ref{sec:knotpress},
small (from
10 per cent in the optical to 25 per cent in the far ultraviolet)
encircled energy
corrections were applied using the information from
www.stsci.edu/hst/stis/documents/handbooks/.  The data were corrected
for reddening due to dust in the line of sight in our Galaxy using the
Galactic extinction curve of Seaton (1979), together with the
expression given in Wilkes et al. (1994) 
(based on Burstein \& Heiles 1978) to convert from $N_{\rm H}$
(see section \ref{sec:spectra}) to $E(B-V)$.

\section{Results}
\label{sec:results}

\subsection{X-ray, optical and radio morphology}
\label{sec:image}

\begin{figure}
\epsfxsize 8.5cm
\epsfbox{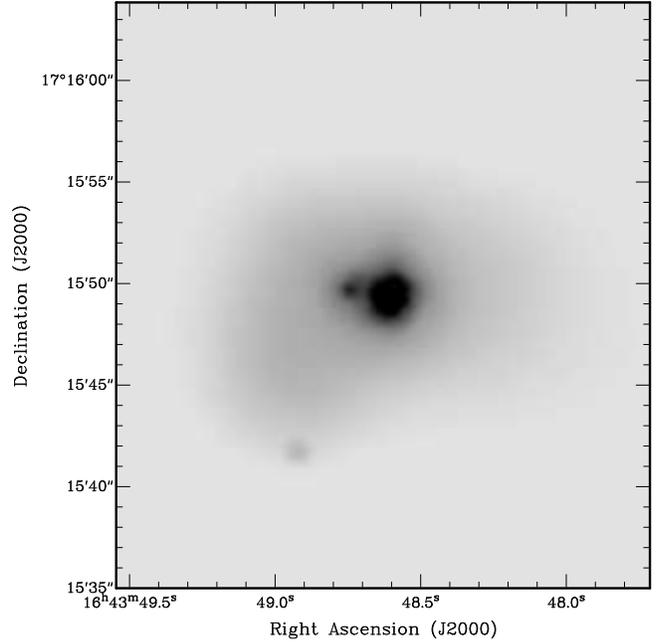}
\caption{
\chandra\ image of \source\ made from adaptively smoothing the
0.3--5-keV data. 
The faint source to the south is unassociated with the
radio or optical structures discussed in this paper.
} 
\label{fig:adaptsmo}
\end{figure}

The image shown in Fig.~\ref{fig:adaptsmo} is the result of adaptively
smoothing the \chandra\ data in the energy band 0.3-5 keV.  It shows the main
X-ray emission components which are

\begin{itemize}

\item a diffuse atmosphere, 

\item bright emission corresponding to the nucleus of \source, and

\item fainter emission 1.95 arcsec east of the core, 
roughly corresponding to the brightest knot of \source's radio and optical jet.

\end{itemize}

\begin{figure}
\epsfxsize 8.5cm
\epsfbox{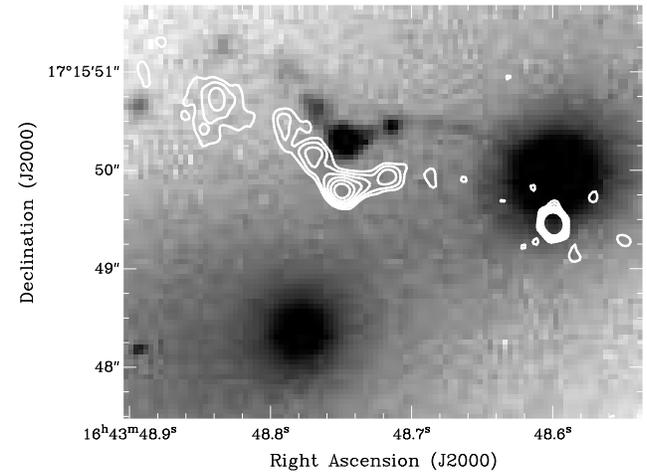}
\caption{
The radio contours of Fig.~\ref{fig:radxray} shown on an
\hst\ archive image (o61p01010 from HST observing program 8786)
taken with the STIS CCD
camera and a clear filter (see section~\ref{sec:opticalobs}).  
The optical data are
shifted north by 0.5 arcsec so that the jet structure can be
seen more clearly.
} 
\label{fig:radoptical}
\end{figure}

The diffuse X-ray emission is not oriented preferentially along the
lobes of the radio galaxy, which span about 17 arcsec to the southwest
and northeast (Spencer et al.~1991; van Breugel et al.~1992; Harvanek
\& Stocke 2002).  Instead, the asymmetry is apparently due to
X-ray-emitting gas encompassing both the host galaxy of \source\ and
its companion galaxy.  Optical emission from the two galaxies and the
jet are shown in Fig.~\ref{fig:radoptical}, where contours from our
high-resolution radio map (section \ref{sec:radioobs}) are also shown,
shifted so that the jet structure can be seen more clearly.  The jet
appears to be almost straight from the VLBI scale (Cotton et al.~1995)
up to the point at which is can be seen above the galaxy emission in
the optical in Fig.~\ref{fig:radoptical}, after which it bends to the
bright radio, optical and X-ray knot (knot C, following the notation
of Dey \& van Breugel 1994), then kinks by about 70 degrees, and
continues to the eastern radio lobe with some further curvature.

\begin{figure}
\epsfxsize 8.5cm
\epsfbox{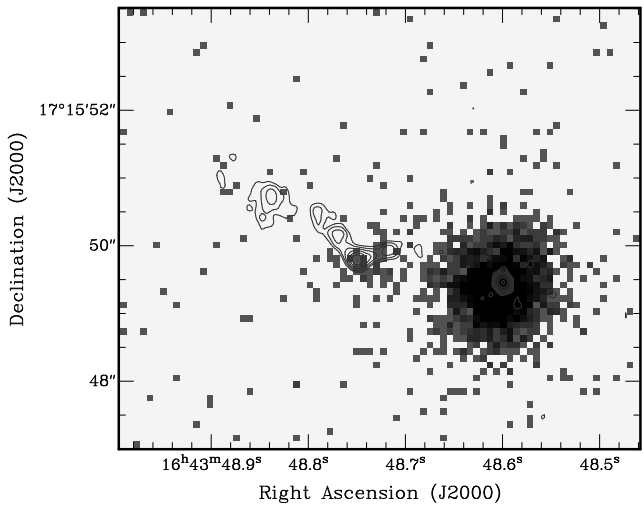}
\caption{
\chandra\ image of the 0.3--5-keV data with no smoothing.
The pixel size is 0.123 arcsec.  15-GHz VLA A-array contours are
shown in grey.  The radio beam size is  0.13 arcsec by 0.11 arcsec.  The
lowest contour is at 0.7 mJy and contour levels increase by factors of 2.
The peak intensity in the core is 206.4 mJy/beam. The images are not
shifted for the 0.12 arcsec offset in core positions that is described more
fully in the text.
} 
\label{fig:radxray}
\end{figure}

\begin{figure}
\epsfxsize 8.5cm
\epsfbox{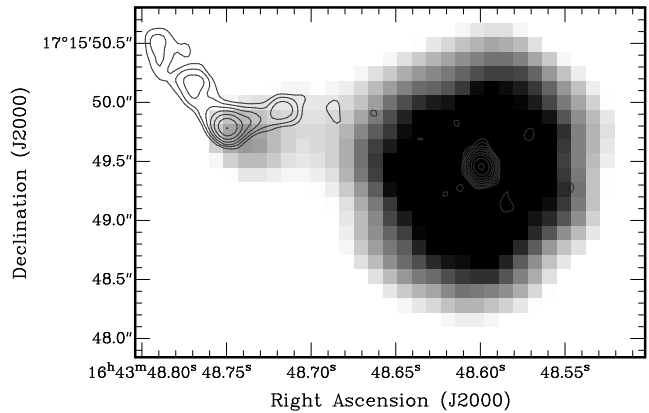}
\caption{
An overlay of the radio contours of Fig.~\ref{fig:radxray} on the
smoothed X-ray image of Fig.~\ref{fig:adaptsmo} finds that
the X-ray peaks slightly closer to the core than the
brightest radio knot. The images are not
shifted for the 0.12 arcsec offset in core positions that is described more
fully in the text.
} 
\label{fig:radxraysmo}
\end{figure}

Figures \ref{fig:radxray} and \ref{fig:radxraysmo} show the radio
contours overlaid on unsmoothed and smoothed \chandra\ data, respectively.  
The jet-related X-rays peak slightly closer to the core than the radio
(and optical) emission.  The excellent spatial agreement between radio and
optical jet features (Fig.~\ref{fig:radoptical}) is already reported
by de Koff et al.~(1996) and de Vries et al.~(1997).

\begin{figure}
\epsfxsize 6.5cm
\epsfbox{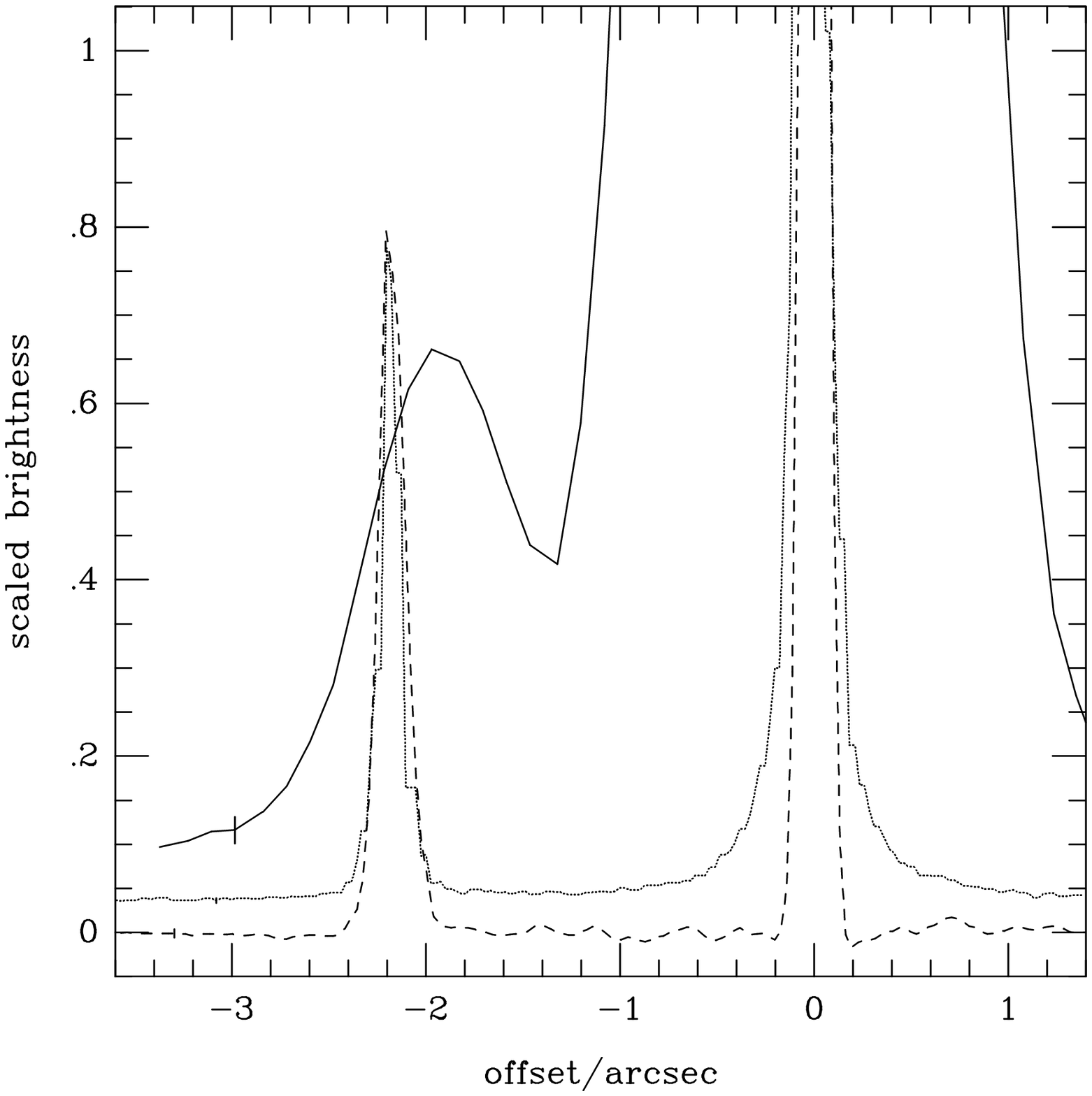}
\epsfxsize 6.5cm
\epsfbox{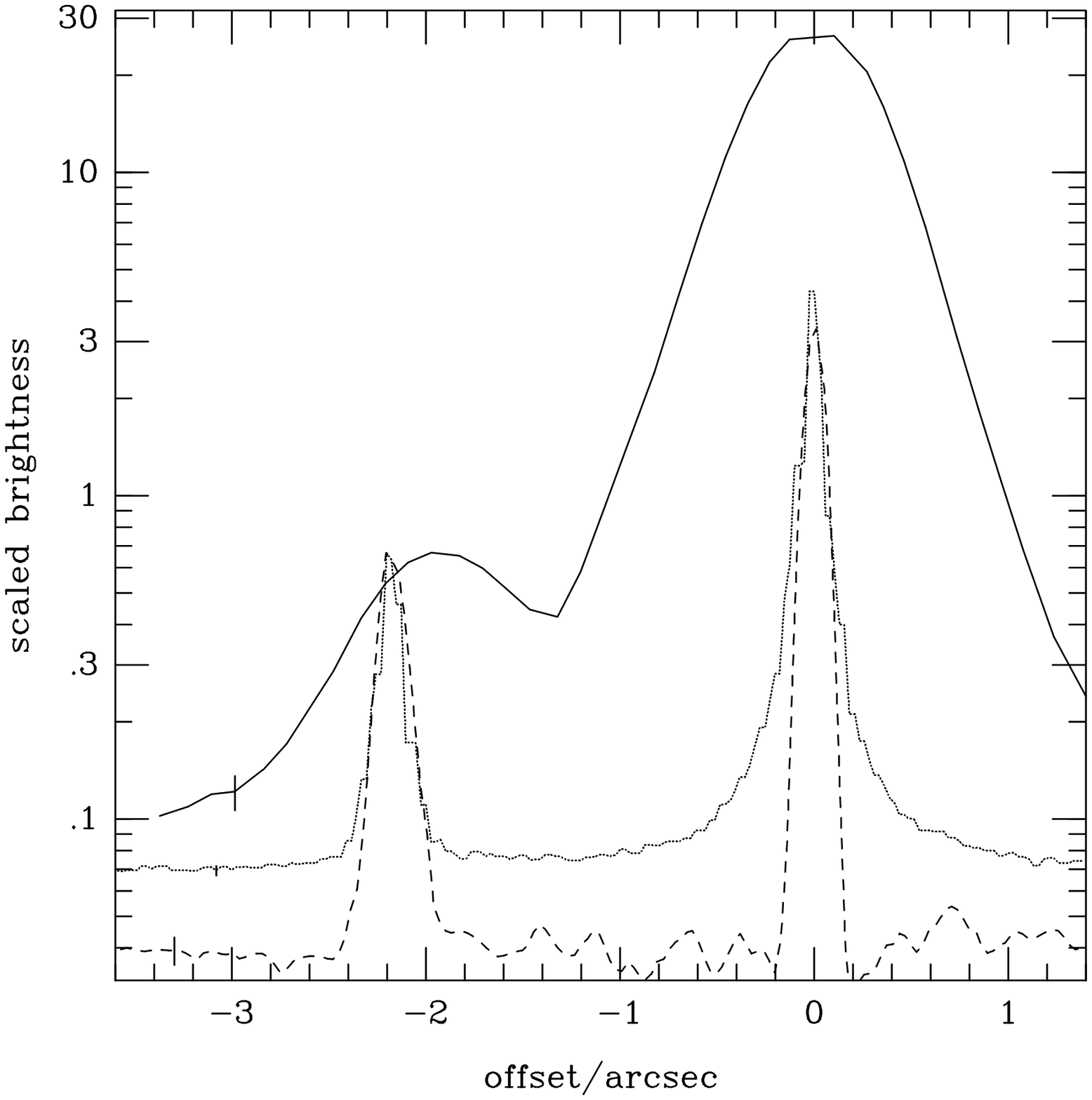}
\caption{
1-D cuts across Gaussian-smoothed ($\sigma = 0.18''$) X-ray (solid
line), unsmoothed \hst\ (dotted line), and
\vla\ 15-GHz (dashed line) images of \source\ along a line in position angle
81.4 degrees connecting the core (at the origin of coordinates) to
knot C. The normalizations are adjusted to show the knot at similar
heights in all three profiles.
Representative error bars are shown at the left of
each profile. The core is located to
better than 0.02 arcsec (radio and optical), 
and better than 0.05 arcsec (X-ray). 
All the cores have been aligned at offset zero.
Note the close alignment of the
radio and optical knots, but the significant ($\sim 0.3$ arcsec) offset of
the X-ray peak.  The upper plot is on a linear-linear scale while the
lower is on a log-linear scale.
} 
\label{fig:slice}
\end{figure}

The radio core in the 15-GHz map is at $\alpha = 16^{\rm h} 43^{\rm m}
48^{\rm s}.599 \pm 0.001$, $\delta = +17^\circ 15' 49''.46 \pm 0.02$.
The centroid of the X-ray core, measured using the {\sc zhtools}
software (A. Viklinin, private communication), is at $\alpha = 16^{\rm
h} 43^{\rm m} 48^{\rm s}.604 \pm 0.003$, $\delta = +17^\circ 15'
49''.36 \pm 0.05$, where the quoted uncertainties are relative to the
astrometric solution derived from the positions of standard stars as
measured by the \chandra\ Aspect Camera.  The images in
Figs.~\ref{fig:radxray} and \ref{fig:radxraysmo} are not shifted for
the 0.12 arcsec offset in these core positions, which is mostly in
declination, since this is within the \chandra\ absolute aspect
uncertainties.  The core and knot in the radio map are separated by
0.35 arcsec more than in the X-ray map, mostly in right ascension.
The largest uncertainty in these measurements is in the X-ray centroid
of the knot, but with 70 counts (see section \ref{sec:spectraknot})
the centroid is located to better than 0.1 arcsec, and therefore we
conclude that the offset between the X-ray and radio knots (a
projected linear distance of 0.97 kpc) is real.  One-dimensional
profiles in X-ray, optical and radio along a line joining the core and
knot C (Fig.~\ref{fig:slice}) also illustrate this result.
When the X-ray profile of the core is scaled to the counts in the knot
and shifted by the 2 arcsec needed for the two to coincide, we find
a good match.  Since the X-ray core is dominated by unresolved emission
(section~\ref{sec:spectracore}) we can conclude 
that there is no indication of X-ray extension in the knot.

\subsection{X-ray radial-profile analysis}
\label{sec:radial}

Fig.~\ref{fig:adaptsmo} shows an envelope of emission extending more
than 10 arcsec (28~kpc).  The wings of the Point Spread Function (PSF) of
the bright core are mixed with this emission, and so we use a
radial-profile analysis to separate the core and galaxy-scale gas
components that were confused in the larger beam of \rosat\ (Worrall \&
Birkinshaw 2001).

\begin{figure}
\epsfxsize 8.0cm
\epsfbox{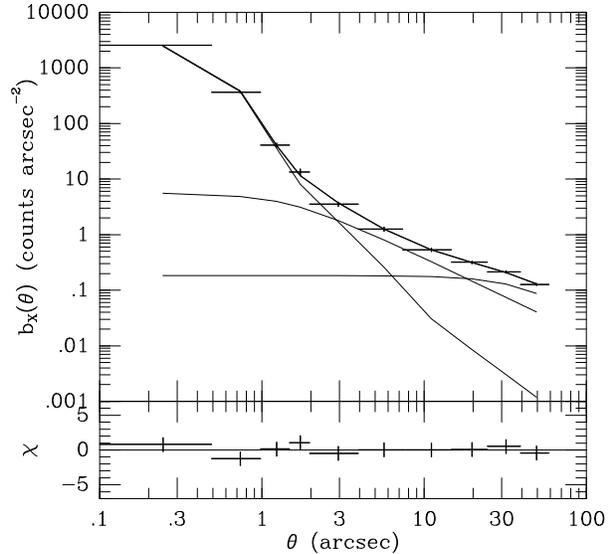}
\caption{
The data points show the background-subtracted 0.3-5 keV radial
profile.  The model is the composite of a point source, a galaxy-scale
$\beta$-model of fitted $\beta = 0.4$, $\theta_{\rm c}$ = 1.3 arcsec,
and a cluster-scale $\beta$-model of fixed $\beta = 0.9$, $\theta_{\rm
c}$ = 78 arcsec.  The $\beta$ models are convolved with the
PSF. Normalization is a fitted parameter for all three
components. $\chi^2 = 4$ for 5 degrees of freedom.
} 
\label{fig:profile}
\end{figure}

Our method for modeling the PSF and fitting models to
background-subtracted data follows Worrall, Birkinshaw \& Hardcastle
(2003).  For extended emission we use a $\beta$ model, which describes
the surface-density distribution with angular radius, $\theta$, for
isothermal gas in hydrostatic equilibrium as proportional to $[1 +
(\theta/\theta_{\rm c})^2]^{0.5 - 3\beta}$ (Cavaliere \& Fusco-Femiano
1978).  The background was measured to be $0.080 \pm 0.003$ cts
arcsec$^{-2}$ (0.3--5 keV) from a rectangular region more than 3
arcmin from \source, and thus beyond the cluster-scale gas component
measured with \rosat\ (Worrall \& Birkinshaw 2001).  This cluster gas
should show up as a roughly flat component in our \chandra\ radial
profile.  We have modeled it as a $\beta$ model, fixing the
parameters to the \rosat-derived best-fit values of $\beta$ = 0.9,
$\theta_{\rm c}$ = 78 arcsec, and allowing the normalization to be a
free parameter.  Both a point source and a $\beta$ model of galaxy
scale must be added to the cluster-scale component to obtain
a good fit to the data (Fig.~\ref{fig:profile}).
Uncertainties in model parameters are determined as described in
Worrall \& Birkinshaw (2001).

\begin{table}
\caption{Radial-profile analysis, 0.3--5 keV}
\label{tab:radial}
\begin{tabular}{ll}
Parameter & Value \\
point-source counts & $3100^{+60}_{-110} $ \\
$\beta$ (galaxy) & $ 0.4^{+0.17}_{-0.07} $ \\
$\theta_{\rm cx}$ (galaxy) & $1.3^{+3.8}_{-1.3} $ arcsec \\
galaxy-model counts, $\theta \leq 10$ arcsec & $ 282^{+72}_{-52} $ \\
$\beta$ (cluster) & 0.9 (fixed) \\
$\theta_{\rm cx}$ (cluster) & 78 arcsec (fixed) \\
cluster-model counts, $\theta \leq 1$ arcmin & $1200^{+550}_{-650}$ \\
cluster-model counts, $\theta \leq 3$ arcmin & $2500^{+1100}_{-1400}$ (extrapolated)\\
\end{tabular}
\medskip
\begin{minipage}{\linewidth}
For the point source and galaxy-scale $\beta$ model, the uncertainties
are $1\sigma$ for two interesting parameters due to the correlation
between $\beta$ and $\theta_{\rm c}$, and between counts in the two
components.
The uncertainty in the cluster-model counts is $1 \sigma$ for one
interesting parameter.
\end{minipage}
\end{table}

\subsection{X-ray spectra}
\label{sec:spectra}

Using our image analysis as a guide, we fitted spectral models to the
0.3-10~keV counts from various regions.  In all cases the X-ray spectra are
binned to a minimum of 30 counts per bin, and all spectral fits
include absorption along the line of sight in our Galaxy assuming a
column density of $N_{\rm H} = 5.6 \times 10^{20}$ cm$^{-2}$ (from the
{\sc colden} programme provided by the CXC, using data of Dickey \&
Lockman 1990).

\subsubsection{X-ray core}
\label{sec:spectracore}

\begin{figure}
\epsfxsize 5.0cm
\epsfbox{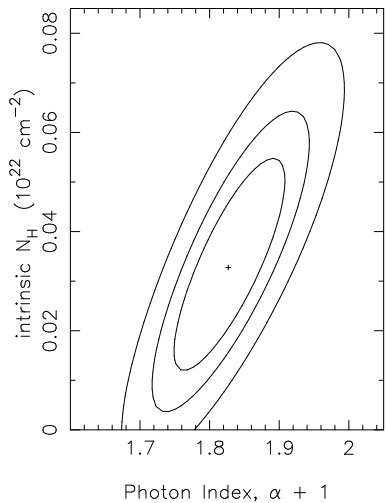}
\caption{$\chi^2$ contours ($1\sigma$, 90\%, and 99\%, for two
interesting
parameters) for a spectral fit to the nuclear X-ray emission extracted
from a circle of radius 1.23 arcsec.  The model is a single-component
power
law with intrinsic, and (fixed) Galactic, absorption.  
  $\chi^2_{\rm min} = 76.3$ for 73 degrees of freedom.
} 
\label{fig:corenhspec}
\end{figure}

An unresolved component heavily dominates the emission
at small radii (Fig.~\ref{fig:profile}).  Within 1.23 arcsec of the centre the
contribution from the resolved thermal components should be less than
1\%.  Taking background from an annulus of radii 5 and 25 arcsec, we
find a good fit ($\chi^2_{\rm min} = 76.3$ for 73 degrees of freedom: 
Fig.~\ref{fig:corenhspec}) to a power law of spectral index $\alpha = 0.83 \pm
0.08$ ($S_\nu \propto \nu^{-\alpha}$) and intrinsic absorption
$(3.3 \pm 2.2) \times 10^{20}$ cm$^{-2}$, where uncertainties are
$1\sigma$ for 2 interesting parameters.
The 1~keV normalization, corrected for the small fraction of missing counts
in the wings of the PSF, is $(1.15 \pm 0.07) \times 10^{-4}$ photons
cm$^{-2}$ s$^{-1}$ keV$^{-1}$, which is $76 \pm 5$ nJy.
The 0.3-5 keV luminosity of the emission (before absorption) is
$3.8 \times 10^{43}$ ergs s$^{-1}$.

If there is an $A_{\rm v}\ga 8$ mag towards the nucleus, as Dey \& van
Breugel (1994) have inferred is necessary to obscure the broad
emission-line region, if one is present, a gas-to-dust ratio similar
to our Galaxy would imply a line-of-sight hydrogen column density to
the nucleus of $\ga 1.4 \times 10^{22}$ cm$^{-2}$.  If there is an
accretion-related X-ray component with the same power-law index as we
have measured lying behind an absorption of $N_{\rm H} = 1.4 \times
10^{22}$ cm$^{-2}$ we can set a 90\% confidence upper limit on its
0.3--5-keV luminosity (before absorption) of $4.2 \times 10^{42}$ ergs
s$^{-1}$.  For a larger $N_{\rm H}$, the luminosity of such a central
emission component could be higher.

\subsubsection{Gas components}
\label{sec:spectragas}

The radial profile shows that the gas component that resembles a
galaxy-scale $\beta$-model is mixed with core
emission at small radii and with cluster-scale gas at large radii.
The annulus in which the galaxy-scale emission should dominate is
rather narrow and will contain significant contributions from these
other components.

Firstly, we extracted the counts in a source-centred circle of radius
9 arcsec, excluding a circle of radius 1 arcsec around the knot to the
east of the core. Background is measured in an annulus of radii 20 and
25 arcsec, so that the contribution of the cluster-scale component
should effectively be removed.  As expected, the fit is dominated by
the slightly-absorbed power-law component that fitted the core
spectrum, but an improved $\chi^2$ (by 12) is obtained when a
thermal component is included.  The fit (which gives $\chi^2$=73
for 85 degrees of freedom) is insensitive to the abundances, which we
fixed at 0.5 solar.  The fitted temperature of the thermal
component is $kT = 0.84 \pm 0.20$ keV ($1\sigma$ error for 2
interesting parameters).  The parameters for the power-law component
are consistent with those found in the fit to the core spectrum.

Secondly, we extracted the counts in a source-centered annulus of
inner radius 5 arcsec and outer radius 1 arcmin.  The counts from an
unrelated X-ray source about 46 arcsec to the south were excluded.
The background was the region more than 3 arcmin from the source that
was used for the radial profile.  We found excellent
fits with Galactic absorption (as might be expected away from the
nucleus, increasing our confidence in the detection of a small
intrinsic absorption for the core), and so no intrinsic absorption is
included in the following results.  The fit to a single-component
thermal of $kT \sim 2.9$ keV and abundances of about 0.5 solar was
formally marginally acceptable ($\chi^2 = 121$ for 98 degrees of
freedom), but two thermal components gave a much improved description
of the data ($\chi^2 = 103$ for 95 degrees of freedom), with the lower
temperature matching that found in the fit to the data from the
9-arcsec-radius circle.  The abundances were not well
constrained and were both set to be 0.5 times solar.  The fitted
temperatures are $kT
= 3.5 \pm 0.6$ keV and $kT = 0.83 \pm 0.15$ keV ($1\sigma$ errors for 2
interesting parameters).

We made further annular extractions, varying the extraction
radii to examine the division of counts between the hot and cold gas
components in the spectral fits, and comparing with expectations
from the radial fit.  When we associated the cooler temperature with the
galaxy-scale component and the hotter temperature with the
cluster, we found agreement with the radial-profile results
as long as the galaxy emission (with flat best-fit $\beta < 0.5$)
truncates at a radius of less than 1 arcmin (about 50 kpc, or 18
arcsec fits best).
We then used the fit of the counts from an annulus of inner radius
20 arcsec and outer radius 1 arcmin to a single-component thermal
to get an improved description of the temperature of the cluster gas,
finding $kT = 2.63 \pm 0.27$ keV ($1\sigma$ error).

Using a combination of the radial and spectral analysis to describe
the distribution of counts with radius and energy, we estimate the
bolometric luminosity of the hot gas in the galaxy to be $(3.0 \pm 0.8)
\times 10^{42}$ ergs s$^{-1}$ out to a radius of 50 kpc, and the total
luminosity of the cluster (integrating out to infinity, since $\beta >
0.5$) to be $(6.5 \pm 2.5) \times 10^{43}$ ergs s$^{-1}$.  The galaxy
atmosphere agrees well with the luminosity-temperature relation for
the atmospheres around other nearby radio galaxies (Worrall \&
Birkinshaw 2000).  The cluster luminosity is marginally lower than
that expected for agreement with the temperature-luminosity
relationship for clusters (e.g., Arnaud \& Evrard 1999), perhaps
suggesting that the cluster is in an interesting dynamical state, but
the uncertainties are large due to the subarray used for the \chandra\
observations and the source's relative proximity, which distributes
cluster counts over much of the image.

\subsubsection{X-ray knot}
\label{sec:spectraknot}

The X-ray counts from the knot were measured using a circle of radius
0.984 arcsec.  Background was measured from five non-overlapping
circles of the same radius at an equal distance from the nucleus.
The knot yielded 70 net counts.  Its spectrum was measured using
the Cash statistic in {\sc xspec} with no intrinsic absorption, and
found to be consistent with $\alpha = 1.0 \pm 0.3$ ($1\sigma$
uncertainty).  The 1 keV flux density is $1.6 \pm 0.2$ nJy.

\section{Discussion}
\label{sec:discussion}

\subsection{Gas properties}
\label{sec:discgas}

There are two X-ray emitting gaseous atmospheres around \source.  The
first, of cluster dimension was known previously, and its temperature
($kT = 2.63 \pm 0.27$~keV) agrees with the \asca\ results of
Worrall \& Birkinshaw (2001), to within the rather larger \asca\
uncertainties. The second, of
galaxy scale and $kT = 0.83 \pm 0.15$~keV is found only because
of the high spatial resolution of
\chandra.   At 2 arcsec from the core, using the
equations of Birkinshaw \& Worrall (1993) to deproject the surface
brightness [see also Worrall \& Birkinshaw (2004) for a more extensive
treatment], we find the pressures of the galaxy gas and cluster gas to
be $7^{+1}_{-3} \times 10^{-12}$ and $(2 \pm 1) \times
10^{-12}$ Pa, respectively. (Uncertainties are 1 sigma for 2
interesting parameters, and these values should be multiplied by 10 to
give units of dynes cm$^{-2}$ used by many authors.)  It is
interesting to note that the radius, $6.3^{+13.7}_{-1.3}$ arcsec,
at which the galaxy-gas and
cluster-gas pressures become equal, and total
$\sim 4 \times 10^{-12}$ Pa, is similar
to the maximum radial extent of the radio source.
Perhaps more interestingly, if the jet were at an angle of $18^{+5}_{-12.4}$ degrees
to the line of sight, the pressures in the two gas
components would be equal at the X-ray-bright knot, the position at which the
radio jet changes direction by almost 70 degrees as seen in projection
(a smaller bend angle being accommodated if the jet is indeed at less than 90
degrees to the line of sight.)
Such an angle to the line of sight is consistent with Giovannini et
al.~(2001; see also Cotton et al. 1995)
who argue, based on jet-sidedness and core dominance,
that the angle to the
line of sight is less than 30 degrees and the bulk Lorentz factor is
greater than 1.7.  We suggest below that the jet bending is the result
of a oblique shock due to interactions between the jet and the wake in
the cluster medium behind \source's companion galaxy.  In this case
the knot must lie beyond the influence of the galaxy gas.
The wake will slightly disturb the galaxy gas from the spherical distribution
we have adopted.  Since there are reasons that the angle to the line of sight
should not be too small (see below), we adopt 20 degrees as the angle
between the line of sight and the jet.  This is consistent with
all the evidence.

Whereas the cooling time of the cluster gas is larger than the
Hubble time, the galaxy-scale gas has a cooling time less than
$10^{10}$ years even out to its maximum radius of 18 arcsec.
Although the gas extends beyond the radio lobes, the radio source should
be providing heat to the gas through the conversion of mechanical
energy. Whether or not this is sufficient to regulate a cooling flow
in the gas is a matter of speculation for this source, whose relatively
high redshift results in the galaxy gas having too little contrast
against the core for temperature structure to be studied.

\subsection{Internal and external pressures at the jet knot}
\label{sec:knotpress}

We measure the bright radio knot (knot C in the notation
of Dey \& van Breugel 1994) to have a 15-GHz flux density
of $65.6 \pm 0.9$ mJy and have a (slightly elliptical)
deconvolved size corresponding to a sphere of radius 0.043 arcsec.
If we assume that the knot is unbeamed and
has equal energy density in radiating particles and magnetic field,
and thus is radiating at close to minimum total energy (assuming no
non-radiating particles and a uniform filling factor), then
the magnetic field is $\sim 72$~nT (720 $\mu$Gauss) and
the pressure in the knot
is $\sim 1.4 \times 10^{-9}$ Pa, which is much larger than the
pressure in the external medium which is no more than $\sim 4 \times
10^{-12}$ Pa.

A way to reduce the internal pressure is to assume that the knot has
bulk relativistic motion towards the observer, in which case under
equipartition the magnetic-field strength is reduced by a factor of
$\delta$ and the internal pressure is reduced by a factor of
$\delta^2$ (e.g., Worrall \& Birkinshaw 2004), where $\delta$ is the
bulk relativistic Doppler factor.  We would require $\delta \sim
19$\footnote{Actually a little larger since with large $\delta$ and
small angle to the line of sight the knot will be further out in the
atmosphere where the external pressure is lower.}  to bring the two
pressures into agreement under these simple assumptions, with the
required values for the components immediately to the east and west of
knot C being $\delta =$ 6.8 and 5, respectively.  A value of $\delta$
as large as 19 is only obtained if the jet is less than 3 degrees to
the line of sight, which we consider unlikely based on the
radio/optical/X-ray energy distribution of the core being similar to
radio galaxies rather than quasars (see Worrall and Birkinshaw 2001),
the absence of broad H$\beta$ (Dey \& van Breugel 1994), and resulting
disagreement with the core-prominence analysis of Cotton et
al.~(1995).

It seems more reasonable to suggest that the
angle to the line of sight is about 20 degrees, and the maximum
relativistic Doppler factor, $\delta$, is of order
3.  In this case, the X-ray-bright knot is overpressured by a factor
of about 40 with respect to the external medium.  
Similar overpressures are seen between terminal hotspots and external
gas (e.g., in Cygnus~A; Carilli et al.~1996), and thus the presence of
an overpressure of the
magnitude we infer in knot C of \source\ is not a proof that the knot
structure is due to something other than a jet fluid/external medium
interaction.
Overpressure is also consistent with the
observation, based on the radio and optical, that the knots
exterior to knot C are considerably larger than those up to the 70-degree bend,
and the inference that the jet expands under the influence of
overpressure at knot C.

\subsection{Jet emission mechanism}
\label{sec:discjet}

\begin{figure}
\epsfxsize 6.0cm
\epsfbox{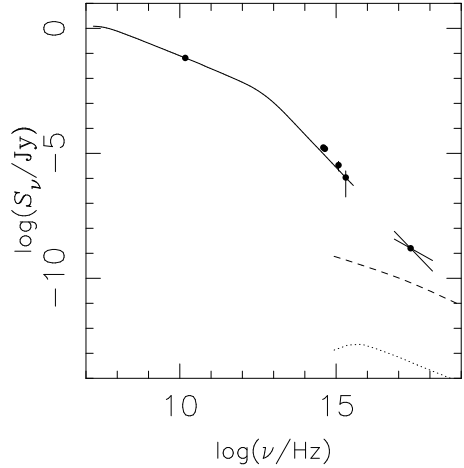}
\caption{
Multiwavelength data for the bright jet knot are compared
with a synchrotron model (full curve) and predictions in an
equipartition magnetic field of
synchrotron self-Compton emission (dashed curve) and inverse Compton
scattering on the cosmic microwave background, assuming no beaming (dotted curve).
The synchrotron spectrum is described by an electron number spectrum
of slope 2.0 breaking by 1.64 at a Lorentz factor of 49000.
} 
\label{fig:synknot}
\end{figure}

The \vla, \hst, and \chandra\ flux densities of the
X-ray-bright knot are given in Table~\ref{tab:knot}, and they are
plotted in Fig.~\ref{fig:synknot}.  The radio spectral index in the
X-ray-bright knot is $\alpha \sim 0.5$, suggestive of particle
re-acceleration (Spencer et al.~1991).

\begin{table}
\caption{Flux densities of the X-ray bright knot}
\label{tab:knot}
\begin{tabular}{ll}
$\nu$ (Hz) & $S_\nu$ \\
$1.50 \times 10^{10}$ & $65.6 \pm 0.9$ mJy\\
$3.90 \times 10^{14}$ & $17.1 \pm 0.9~\mu$Jy \\
$4.37 \times 10^{14}$ & $15.2 \pm 0.3~\mu$Jy \\
$1.19 \times 10^{15}$ & $3.4 \pm 1.4~\mu$Jy \\
$2.03 \times 10^{15}$ & $1.1 \pm 0.9~\mu$Jy \\
$2.4 \times 10^{17}$ & $1.6 \pm 0.2$ nJy \\
\end{tabular}
\end{table}

Using methods described in Hardcastle, Birkinshaw \& Worrall (2001),
we find that synchrotron-self-Compton emission with an equipartition
magnetic field under-predicts the X-ray emission, and any beaming will
reduce it further.  Inverse Compton scattering of the cosmic microwave
background (CMB) is increased if the jet is highly relativistic and at
small angle to the line of sight.  If the magnetic field is held at
72~nT, the observed luminosity at a given frequency from
inverse-Compton scattering of an isotropic radiation field (such as
the CMB) will be increased relative to the synchrotron luminosity by
$\delta^{1+\alpha}$ or, if the field is reduced by a factor of
$\delta$ to bring it back into equipartition, the relative increase is
by $\delta^{2+2\alpha}$ (Dermer 1995).  For the factor of $\sim 10^4$
increase needed to explain the X-rays by this mechanism, we would
require the unlikely situation (see above) that $\delta$ is about 450
or 22, for the two assumptions concerning the magnetic field. We thus
conclude that the jet X-ray emission is synchrotron radiation.

There are two features that are in common with results reported
previously for closer synchrotron jets.  The most obvious, seen in
Figs.~\ref{fig:radxraysmo} and \ref{fig:slice}, is the offset between
the radio and X-ray emission, with the X-ray emission arising from
closer to the nucleus. (The optical and radio knots show strong
correspondence.)  This phenomenon, first
reported in 3C\,66B (Hardcastle et al.~2001), is now seen in a number
of sources, most notably Cen\,A, whose proximity gives the best
spatial resolution and permits interpretation of the data in terms of
a shock model where the acceleration of low-energy electrons that
produce strong radio emission occurs in a turbulent shock region
downstream of a main shock (Hardcastle et al.~2003).  Our
interpretation of \source\ in section \ref{sec:oblique} could produce
a similar situation.

The second feature is the size of the spectral break in a
synchrotron component connecting the radio and optical emission.
The value for \source\ ($\sim 0.8$) is within the range $\Delta \alpha \sim 0.6$ to
$0.9$ seen in other sources, 
with M\,87 (B\"ohringer et al.~2001) and 3C\,66B (Hardcastle et
al.~2001) having some of
the best available multiwavelength coverage.
Synchrotron energy losses within the
context of a simple continuous-injection model predict $\Delta \alpha
= 0.5$.  However, the offset between the X-ray and radio emission
already tells us that there is spatial
substructure, and it is likely that the radio and optical regions
contain a sum over different electron spectra.

The radiative half-life of electrons producing synchrotron radiation of
frequency
$\nu$~Hz in a magnetic field of $B$ Tesla is $\tau \sim 0.043 B^{-3/2}
\nu^{-1/2}$ years, which in the case of \source's X-ray emission is
$4.5 \delta^2$ years in an equipartition magnetic field.
For $\delta = 3$, this implies an X-ray-emitting region which is only about
5 per cent of the measured radio size of knot C unless 
there is distributed particle acceleration, and there is a
chance that a further deep \chandra\ observation might
detect flux variability.

\subsection{The core}
\label{sec:disccore}

\chandra\ has improved upon previous knowledge of the core spectrum
from \rosat\ and \asca\ (Worrall \& Birkinshaw 2001) by turning upper
limits to the intrinsic column density into a detection of $(3.3 \pm
2.2) \times 10^{20}$ cm$^{-2}$ (Section~\ref{sec:spectracore},
Fig.~\ref{fig:corenhspec}).  The relatively small intrinsic absorption
measured for the X-ray core supports our earlier conclusions that the
emission is jet-related in origin.  The \chandra\ spectral index is
consistent with that found from \rosat\ and \asca, and the fact that
the X-ray emission from the galaxy is only about 10\% of the strength
of the core makes it unlikely that a varying degree of confusion of
galaxy emission in the different beam sizes of \rosat\ and \asca\ is
responsible for the reported X-ray variability of $\sim$ 32\% between
\rosat\ and \asca\ observations.  The \chandra\ normalization is in
broad agreement with that found with \asca.

\section{Interpretation of the jet}
\label{sec:interpretation}

The basic question to ask of the X-ray bright knot (as of the other
knots in the jet) is whether it has come from the nucleus with roughly
its current properties, or whether there is some physical process at
its present location which has made it so bright. That is, is the jet
largely a \textit{ballistic} or \textit{hydrodynamic} phenomenon?

\subsection{Ballistic jet?}
\label{sec:ballistic}

One can argue for the ballistic case on the basis of the presence of a
number of knots within the continuous optical structure that extends
from the centre of \source's host galaxy to knot~C and beyond. The
increase in brightness of the optical knots as knot~C is approached,
and then the decrease after that point, might be taken as evidence
that knot~C is bright because of a combination of geometry and
relativistic effects.

It is possible to construct a kinematic model for the shape of the jet
of \source\ based on a precessing beam of ballistic knots,
although it is necessary to allow for some decrease in the speed of
the knots as they propagate. Knot~C and its associated sharp bend
can then be interpreted as a cusp in the projected figure
of motion of components propagating away from the
core. Under this interpretation, knot~C is bright because of a
combination of a long line-of-sight path through the jet and
relativistic beaming. If 
this is the case, we would expect the structure of knot~C to be 
complicated when viewed at high angular resolution, and we would also
expect the knot to show low radio and optical polarization, because 
several independent regions are being seen in superposition. Thus the
high (17 per cent) radio polarization of knot~C measured at 8.4~GHz by
Akujor \& Garrington (1995) argues against a ballistic model for the
jet. Furthermore, there is no simple explanation for the offsets
between the radio and X-ray peaks of knot C in this model.

Another reason why we disfavour a ballistic model is based on the
changing angular sizes of the knots to either side of knot~C. 
In a ballistic model, the apparent
expansion of knots between $1$~and $3$~arcsec
from the core implies a real increase in the proper sizes
of the knots.  Adiabatic losses should therefore
cause the synchrotron emissivity of the knots to drop by a factor of
10 or more between the knots west of knot C and those to the east, while
the relativistic boosting factors of knots west and east should
be similar. It is
apparent from Fig.~\ref{fig:radoptical} that the
total fluxes in the knots immediately west and east of knot C are, in
fact, comparable.
The ballistic model is then viable only if the intrinsic
emissivity of the knots increases with distance from the core,
presumably as a result of interaction with the external medium.

Finally, we note that the knots on either side of knot~C have
different morphologies. To the west the knots are unresolved across
the jet, while to the east they are resolved both across and along the
jet. Assuming steady expansion of ballistic knots, a major change of
knot morphologies across knot~C requires a fine-tuned alignment of the
line of sight with the edge of the cone defined by the velocity
vectors of knots.

\subsection{Hydrodynamic jet?}
\label{sec:hydro}

\subsubsection{An oblique shock as an explanation for the kink at knot C}
\label{sec:oblique}

The abrupt change of direction of the jet at knot~C can alternatively
be attributed
to an oblique jet shock. While the $\sim 70$~degree change in
direction of the jet at knot~C is larger than the maximum change of
direction of a flow in an oblique shock ($49^\circ$ for a relativistic
fluid), relatively
modest orientation effects (such as the jet lying about 20~degrees to
the line of sight; section~\ref{sec:discgas}) are sufficient to make
the projected bend at knot~C as large as 70~degrees.

The strong X-ray and radio emission from knot~C can then be
interpreted as the result of enhanced particle acceleration near the
jet shock. The X-rays would be expected to come from an unresolved
component, since the lifetime of X-ray emitting electrons is at most a
few tens of years (section~\ref{sec:discjet}). The radio-emitting
electrons are of significantly lower energy and longer radiative
lifetimes, and so might both be accelerated and be detectable over a
larger volume, perhaps out to a weaker shock, beyond the jet, caused
by the supersonic advance of knot~C into the intergalactic
medium. This might explain the offset between the radio and X-ray
centroids of the knot (Fig.~\ref{fig:radxraysmo}).

The oblique-shock model would suggest that the jet flow remains
supersonic after it
passes through the jet shock, although it is slowed and broadened
there. The obliqueness of the jet shock can also allow knot~C to advance
relatively slowly into the intergalactic medium, so that the
downstream segment of jet, to the E of knot~C in
Fig.~\ref{fig:radoptical}, can form.

\begin{figure}
\epsfxsize 8.5cm
\epsfbox{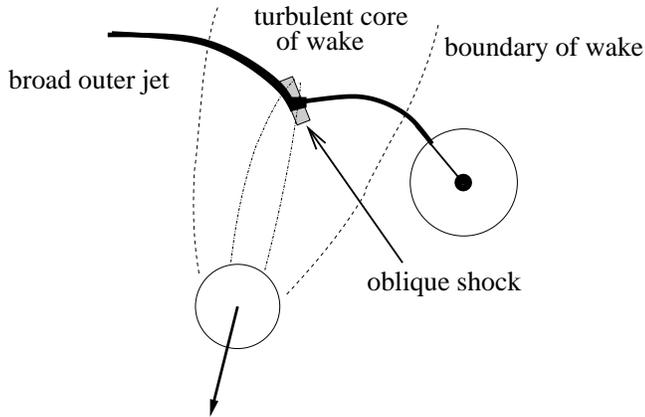}
\caption{
A sketch showing the formation of a wake due to the motion of the
companion galaxy in the cluster and the subsequent bending
and brightening of the jet.  The X-ray emission is assumed to come
from the region of the oblique shock.
} 
\label{fig:sketch}
\end{figure}

It is known (Dey \& van Breughel 1994) that the relative line-of-sight
velocity between the companion galaxy and the host of \source\ is $470
\pm 80 \ \rm km \, s^{-1}$, or about half the speed of sound in the
intergalactic gas. This suggests that the wake of the companion is
likely to be strong, and so knot~C might occur where the jet of
\source\ crosses the turbulent centre of the wake.
Fig.~\ref{fig:sketch} sketches the proposed geometry.

While the idea of an oblique shock is speculative, it could be investigated by
high-sensitivity mapping of the temperature and velocity widths of
X-ray lines from the gas near the jet, and a detailed study of the
radio and optical morphologies of knot~C.

Under this interpretation, we predict that the oblique shock will
generate a relatively simple polarization pattern in the brightest
part of knot~C, with the magnetic-field direction at position angle
$\sim 20$~degrees. This is consistent with the high radio polarization
measured by Akujor \& Garrington (1995), but improved
high-resolution radio and optical polarization studies of \source\
are needed to test this prediction, and so to distinguish between
the oblique-shock and ballistic models.

\subsubsection{General jet curvature}
\label{sec:bending}

If we adopt the concept of an oblique jet shock, then the jet is
interacting strongly with the medium through which it moves
supersonically, 
and we
should explain the curvature in the jet as a hydrodynamic phenomenon.
We note that the curvature is concave upwards on either side of
knot~C. We interpret this as due to a large-scale pressure gradient.

If a pressure gradient is to cause bending on a projected scale $R
\sim 6$ kpc in the cluster-scale atmosphere where the pressure
gradient is low (since the core radius $r_c \sim 220$ kpc $\gg R$), 
the jet must be light and highly supersonic.  Under
these conditions the jet opening angle will be small, as is observed
(Fig.~\ref{fig:radoptical}), and the bending of the jet would be accompanied by a
number of internal and external shocks as the jet responds to the
external pressure gradient.  A calculation assuming that these shocks
are weak, that the jet responds as a homogeneous cyclinder to the
external pressure gradient, and that the flow is steady and
non-relativistic then leads to an estimate of the jet density
\begin{equation}
  \rho_{\rm jet} \approx 
                 \rho_{\rm ext} \, {R \over r_{\rm c}} \,
                  {1 \over {\cal M}_{\rm jet}^2} 
\end{equation}
where ${\cal M}_{\rm jet}$ is the Mach number and
$\rho_{\rm ext} \sim 5 \times 10^{-24}$ kg m$^{-3}$ is the density of
the cluster gas near the jet.  If we also assume that the kinetic
energy flux down the jet is sufficient to power the eastern lobe of
the radio source, we find that the jet speed must be relativistic, and
beaming effects sufficient to make the source appear one-sided are
likely for the jet orientation suggested in section~\ref{sec:discgas}.  Under
these circumstances, detailed calculations of the characteristics of
the flow are necessary (and equation 1 overestimates the density of
the jet).

This bending model can be tested by looking at the polarization
pattern of the inner jet knots, where we would expect to see ordering
of the magnetic field by fans of weak shocks, or by detailed X-ray
spectroscopy of gas near the jet, where we might expect to see local
heating.  While the former test should be possible by detailed 
HST or VLA polarimetry, the necessary X-ray spectroscopy is currently
impractical.

\section{Summary}
\label{sec:summary}

The \chandra\ observations have detected two new X-ray components
associated with \source.  The first is emission from the bright knot
in the jet, and the second is a relatively luminous X-ray-emitting
atmosphere of galaxy scale.  The knot X-ray emission is too bright for
inverse Compton models, and we infer that the emission mechanism is
synchrotron, and the region is one of particle re-acceleration.  There
are similarities to emission regions in the inner regions of low-power
FRI jets both in the multiwavelength spectrum and the fact that the
X-ray emission peaks closer to the core than the lower-frequency
emission.  The radio jet bends by an apparently large angle of about
70 degrees (less if, as we favour, the jet is at about 20 degrees to
the line of sight) and continues.  Thus the knot in \source\ does not
appear to have the nature of a terminal hotspot, for which there are
some cases where synchrotron radiation is the favoured X-ray emission
mechanism (e.g., Harris, Leighly \& Leahy 1998; Wilson, Young \&
Shopbell 2001), but rather a re-acceleration region after which the
jet re-collimates and continues.  It is therefore interesting that we
appear to be seeing in \source\ a more powerful version of phenomena
observed in nearby, low-power jets.

We have suggested that an oblique shock associated with interactions
between the jet and the wake in the cluster medium behind \source's
companion galaxy is responsible for the X-ray-bright knot and the
sharp bend in the jet.  Pressure gradients in the
cluster atmosphere can cause the steady
curvature in the jet each side of the knot.  Since the knot must then
lie beyond the influence of the galaxy atmosphere, the angle to the
line of sight must be about 20 degrees or less, although angles
smaller than about 20 degrees are disfavoured based on the fact that
\source\ differs observationally from a quasar in several ways.

High-resolution studies of the polarization properties of the jet
(Perlman et al., in preparation)
in comparison with the multiwavelength data should
help to describe the details of the acceleration structures and
mechanisms, and should test the hydrodynamic model we favour against a
ballistic model.
Velocities for more members of the cluster would test the hypothesis
that it is \source's host galaxy that is at rest and the companion
galaxy that is moving with respect to the cluster gas.

\section*{Acknowledgements}

We than the CXC for its support of \chandra\ observations,
calibrations, and data processing.
This work also used observations made with the NASA/ESA Hubble Space
Telescope, obtained from the data archive at the Space Telescope
Science Institute, and the VLA archive supported by the
National Radio Astronomy Observatory.
The CXC is operated for NASA by the Smithsonian Astrophysical
Observatory, STScI is operated by the Association
of Universities for Research in Astronomy, Inc. under contract from NASA,
and the NRAO is a facility of the National Science Foundation operated under
cooperative agreement by Associated Universities, Inc.
We also thank the anonymous referee for helpful suggestions as to
where the clarity of the original text could be improved.

\label{lastpage}

\end{document}